\documentclass[allclo]{FBSart}
\usepackage{amsfonts}
\usepackage{amssymb}
\usepackage{amsfonts}
\usepackage{amsmath}
\usepackage{amssymb}
\usepackage{amsthm}
\usepackage{epsf}
\usepackage[dvips]{graphicx}
\usepackage{cite}

\def\bp{{\mbox{\boldmath$p$}}}
\def\bgam{{\mbox{\boldmath$\gamma$}}}

\runningtitle{Style File for Few-Body Systems}
\runningauthor{S.\,M.\ Dorkin et al.} 
\begin{document}

\title{Heavy pseudoscalar mesons in a Schwinger-Dyson--Bethe-Salpeter
approach\thanks{Based on materials of the contribution
"Relativistic Description of Two- and Three-Body Systems in Nuclear Physics",
 ECT*, October 19-23, 2009}}
\author {S.M. Dorkin\instnr{1,2}, T. Hilger\instnr{3,5},
L.P. Kaptari\instnr{1,4},
 B. K\"ampfer\instnr{3,5}} \instlist{
 Bogoliubov Lab. Theor. Phys. JINR, 141980  Dubna, Moscow reg., Russia
\and International University Dubna, Dubna, Russia \and TU Dresden,
Institut f\"ur Theoretische Physik, 01062 Dresden\and Dept. of
Phys., Univ. of Perugia and
      INFN, Sez. di Perugia,
      via A. Pascoli, I-06123, Italy; Supported through the program Rientro dei
Cervelli of the Italian Ministry of University and
Research
\and Forschungszentrum Dresden-Rossendorf, PF 510119, 01314
Dresden, Germany}
\date{Received: date / Accepted: date}

\maketitle

\begin{abstract}
 The mass spectrum of heavy pseudoscalar mesons, described as quark-antiquark bound systems,
 is considered within the  Bethe-Salpeter formalism with momentum-dependent masses of the constituents.
 This dependence is  found by  solving   the Schwinger-Dyson equation for quark propagators
 in rainbow-ladder approximation. Such an approximation is known to provide both a
 fast convergence of  numerical methods
 and accurate results for lightest mesons. However, as the meson mass increases, the method becomes less stable
 and special attention must be devoted to details of numerical means of solving the corresponding equations.
 We focus on the pseudoscalar sector and show that our numerical scheme describes fairly accurately the $\pi$, $K$, $D$, $D_s$
 and $\eta_c$ ground states. Excited states are considered as well. Our calculations are
 directly related to the future physics programme at FAIR.
\end{abstract}

\section{Introduction}

 The description of mesons as quark-antiquark bound states within the framework of the Bethe-Salpeter equation
 with momentum dependent quark masses, determined by the Schwinger-Dyson equation, is able to explain
 successfully many spectroscopic data \cite {JM-1,JM-2,rob-1,rob-2,tandy1,David,Alkofer,fisher,Roberts}.
 Contrarily to traditional phenomenological models, like   quark bag  models,
 the presented formalism  maintains   important features of QCD, such as dynamical chiral
 symmetry breaking, dynamical quark dressing, requirements of the renormalization group theory etc.
 The main ingredients here are  the full quark-gluon vertex function and  the dressed gluon
 propagator, the calculation of which is entirely determined  by  the running coupling
 and the bare quark masses. In principle, if one were able to solve the Schwinger-Dyson equation within all pQCD orders,
 the approach would  not depend on  any free parameters. However,due to  known technical problems,
 one restricts oneself to  calculations of the few first terms of the perturbative series,
 usually up to  the one-loop approximation. The obtained results, which formally  obey  all the
 fundamental requirements of the theory, are then considered as exact ones with, however, effective parameters.
 This is known as the rainbow-ladder approximation for the Schwinger-Dyson equation.
 The merit of the approach is that, once the effective parameters are fixed,
 the whole spectrum of known mesons is supposed to  be described   on the same footing:
 including also excited states.

It should be noted that there exists other approaches based on the same physical ideas
but not so sophisticated, e.g. employing simpler interactions, such as a  separable interaction for the effective
coupling \cite{David}. Such approaches  describe fairly well the properties of light mesons,
nevertheless, investigation of  heavier mesons and excited states,
consisting even of light (u,d,s)  quarks, requires implementations of
more accurate numerical methods to solve the corresponding equations.
Among other   successful   efforts in this realm the Refs. \cite{krassnigg1,souchlas} must be also mentioned.

In the present note we are going to apply the combined Schwinger-Dyson and Bethe-Salpeter (BS) formalisms to
describe the meson mass spectrum including heavy mesons and excited states as well.
Particular attention is paid to the charm sector which, together with the
 baryon spectroscopy, is a major topic in the FAIR research programme.
Two large collaborations at  FAIR \cite{CBM,PANDA} plan precision measurements. Note, that  it becomes now
possible to experimentally investigate  not only the mass spectrum of the mentioned mesons, but also
different processes of their decay, which are directly connected with fundamental QCD problems
(e.g., $U(1)$ axial anomaly, transition form factors etc.) and with the known problem of changing the meson characteristics
at finite temperatures. The latter is  crucial in understanding the di-lepton yields in nucleus-nucleus collisions
at, e.g. HADES.  All these circumstances  require  an adequate theoretical foundation to describe
the meson spectrum and the meson covariant wave functions (i.e. the BS partial amplitudes) needed in   calculations
of the relevant  Feynman diagrams and observables.

Our paper is organized as follows.
In section \ref{dse}, the parametrization of the gluon propagator is
presented and the Schwinger-Dyson equation for the quark propagator is discussed.
Section \ref{s:bse} deals with the Bethe-Salpeter equation.
Numerical results are discussed in Section \ref{num}.
Conclusions are drawn in section \ref{concl}.

\section{Propagators and Schwinger-Dyson equation}
\label{dse}

 The  Bethe-Salpeter and Schwinger-Dyson equations in Minkowski space contain  poles and
 branch-point singularities which strongly hinder the procedure of finding   numerical solutions.
 Usually, to avoid these difficulties, one performs the Wick rotation and formulates the corresponding
 equations in Euclidean space, where all singularities in amplitudes and propagators are removed, so that
 the equations can be solved numerically. The known Mandelstam technique
 allows then to calculate matrix elements of observables which, being analytical functions of the relative energy,
 are the same
 in both  Minkowski and Euclidean spaces.

 In our case we consider  the Schwinger-Dyson equation for the quark propagator within  pQCD
 with summing   all diagrams  up to  one-loop. In calculations of diagrams
 the chiral symmetry breaking is implemented  from the  very beginning \cite{Roberts}.
The exact  results, even  only  up to   one-loop  diagrams, after proper regularization
and renormalization procedures  are rather cumbersome for further
numerical calculations. Nevertheless, in practical calculations one can  employ
reasonable parametrizations for the corresponding vertices and propagators  to find
  solutions numerically.
So, a phenomenological expression, inspired by calculations of the mentioned diagrams
and preserving the requirements of the theory, for  the combined running coupling and
gluon propagator has been suggested in Ref. \cite{Roberts}

\begin{eqnarray}
 g^2(k^2) D_{\mu \nu} (k^2) =
    \left(
        \frac{4\pi^2 D k^2}{\omega^2} e^{-k^2/\omega^2}
        + \frac {8\pi^2 \gamma_m F(k^2)}{
            ln[\tau+(1+\frac{k^2}{\Lambda_{QCD}^2})]}
    \right)
    \left( \delta_{\mu\nu}-\frac {k_{\mu} k_{\nu}}{k^2} \right) \: ,
\label{phenvf}
\end{eqnarray}
where the first term originates from  the infrared (IR) part of the interaction
determined by non-pertubative effects, while the second one ensures the correct
ultraviolet asymptotics. Accordingly to the known fact that
the contribution of the IR part is predominant for formation of bound states,
 in  what follows  we neglect the second term
 and   restrict ourselves to the IR one. Then, as seen from Eq. (\ref{phenvf}), we are left
 with only  two effective parameters, $D$ and $\omega$.

The calculation of the renormalized Feynman diagrams leads to a
fermion propagator depending  on two additional parameters. In  canonical calculations these are
the renormalization constant $Z_2$ and the self-energy corrections $\Sigma(p)$. Usually, for further simplifications
of calculations, instead of $Z_2$ and  $\Sigma(p)$ one introduces other two quantities $A(p)$ and $B(p)$
in terms of which the  quark propagator $ S_q(p)$ reads as
\begin{eqnarray}
    S_q^{-1}(p)= i \gamma \cdot p A(p)+ B(p);  \quad \quad S_q(p) =
    \frac{-i\gamma \cdot p A(p)+ B(p)} {p^2 A^2(p) + B^2 (p)} \: .
    \label{quarkpr}
\end{eqnarray}

Then with such a representation of the quark propagator the   Schwinger-Dyson equation
in  Euclidean space  has the form (cf. Refs. \cite{Alkofer,Roberts})
\begin{eqnarray}
S_q^{-1}(p)= i \gamma \cdot p + \tilde{m}+ \frac 43 \int \frac
{d^4 l }{(2\pi)^4} \left[g^2 D(p-l)\right]_{\mu \nu} \gamma_{\mu} S_q(l)
\gamma_{\nu} \: ,
\label{sde}
\end{eqnarray}
where $\tilde{m}$ is the bare quark mass and the effective kernel $D(p-l)$ is
\begin{equation}
  D(k^2)_{\mu\nu} = \frac{4\pi^2 D k^2}{\omega^2} e^{-k^2/\omega^2}\left( \delta_{\mu\nu}-\frac {k_{\mu} k_{\nu}}{k^2} \right).
\label{kernel}
\end{equation}
 Note that  (\ref{sde}) is a    four dimensional integral equation.
 To solve it one usually decomposes the kernel over a complete set of basis functions,
performs analyticaly  some angular integrations    and considers a new system of equations relative
to such a partial decomposition. In our calculations we
  expand the interaction kernel  into hyperspherical harmonics
\begin{eqnarray}
    D((p-l)^2)= D(p^2,l^2,|p||l|\cos \gamma ) = \sum \limits_n
    D_n(p,l)C^1_n (\cos \gamma) \: ,
    \nonumber \\
    D_n(p,l)= \frac {2}{\pi} \int \limits_{-1}^1 D(p^2,l^2,p\cdot l\cdot t) C^1_n(t)
    \sqrt{1-t^2} dt \: ,
\label{Dexp}
\end{eqnarray}
where $C_n^1(t)$ are the Chebyshev polynomials.
For the employed ansatz of the gluon propagator the angular integration can be performed analytically leading to
\begin{align}
    D_n(p,l) & = (n+1) 8 \pi^2 D e^{-x} \left[ \frac{x-n}{z} I_{n+1}(z) - I_{n+2}(z) \right] \: ,
    \\
    D^\prime_n(p,l) & = (n+1) 4 \pi^2 \frac{D}{\omega^2} e^{-x} \frac{I_{n+1}(z)}{z}
\label{ali}
\end{align}
with $x=(p^2+l^2)/\omega^2$, $z=2pl/\omega^2$ and $I_n(z)$ being the modified Bessel functions of the first kind.
In Eq. (\ref{ali}) $ D^\prime_n(p,l)$ denotes the corresponding part of the kernel $D(k^2)/k^2$.
Eventually, the system of equations to be solved reads as
\begin{eqnarray}
    A(p)&=&1+\frac {4}{3p} \int  \frac {dl l^4}{16\pi^2} \frac
    {A(l)}{l^2A^2(l)+B^2(l)} D_1(p,l)+
    \nonumber \\&&
    \int \frac {dl l^4}{16\pi^2} \frac {A(l)}{l^2A^2+B^2}
    [\frac 83( p + \frac {l^2}{p})D^{'}_1(p,l)-\frac 43 l
    (D_2^{'}(p,l)+5D_0^{'}(p,l))] \: ,
    \nonumber  \\
     B(p)&=&\tilde m + 4
    \int \frac{dl l^3}{8\pi^2} \frac {B(l)}{l^2A^2(l)+B^2(l)} D_0(p,l) \: .
\label{ab}
\end{eqnarray}

\begin{figure}
\begin{center}
    \includegraphics[width=0.8\textwidth]{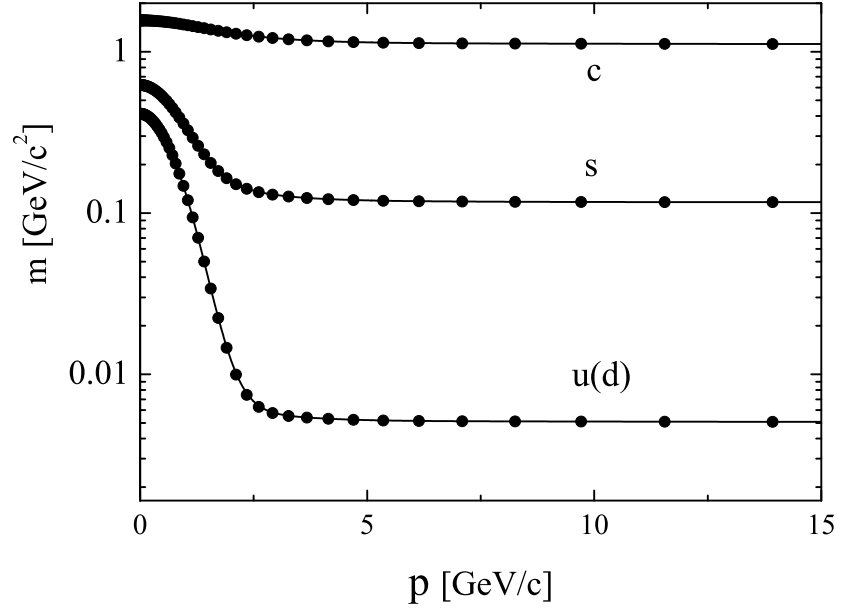}
    \caption{The solution of the Schwinger-Dyson equation for the quark mass $m(p) = B(p) / A(p)$
    as a function of momentum for different flavors, i.e. for the input bare quark masses $\tilde{m}$.
    At asymptotically large  values of $p$ the mass of the dressed quark tends to its  bare value.}
    \label{fig1}
    \end{center}
\end{figure}

The resulting  system of equations on  (\ref{ab}) is a  system of one-dimensional
integrals and  can be solved numerically, e.g. by an iteration method. We found that iteration
procedure for  (\ref{ab})  converges  rather fast and  practically does not depend up on
the choice of the trial functions for $A(p)$ and $B(p)$.
The numerical solution of  (\ref{ab}) with the effective parameters from Ref. \cite{Alkofer,Roberts}
is shown in Fig.~\ref{fig1} as the momentum dependence of the mass $m(p) = B(p) / A(p)$ of the
dressed  quark for different quark flavors, i.e. for different bare masses:
$\tilde{m}=0.005$ GeV for $u(d)$, $\tilde{m} = 0.115$ GeV for $s$
and $\tilde{m}=1$ GeV for $c$ quarks \cite{Alkofer}.
In principle, the asymptotic behavior of $A(p)$ and $B(p)$ for large momenta
can be obtained directly from \eqref{ab} and, as an additional check of the method,
compared with the corresponding numerical results.

\section{Bethe-Salpeter vertex function}
\label{s:bse}

To determine the bound state mass of a quark-antiquark pair one needs to solve the Bethe-Salpeter equation.
In rainbow-ladder approximation and in Euclidean space it reads \cite{Alkofer,Roberts}
\begin{eqnarray}
    \Gamma(P,p) =  (-\frac 43) \int \frac {d^4k}{(2\pi)^4}
    \gamma_{\mu}S(k_+) \Gamma(P,k) S(k_-)\gamma_{\nu} (g^2
    D(p-k))_{\mu \nu} \: ,
\label{bse}
\end{eqnarray}
with $\Gamma$ being the Bethe-Salpeter vertex function.
The color structure has been factorized explicitly.
$P$ means the total momentum of the bound state, $p(k)$ is the relative momentum within the quark pair and $k_+ = k + \xi P$, $k_- = k + (\xi - 1) P$.
The result does not depend on $\xi$ due to the covariance of the Bethe-Salpeter equation.
We choose $\xi = 0.5$.
However, approximations might destroy covariance and one has to check a posteriori the stability of the results.

Equation (\ref{bse}) is written in matrix form, i.e. the vertex
function $\Gamma (P,k)$ is a 4x4 matrix and, therefore, may contain
16 different functions.
The general structure of the  vertex functions
describing bound states of spinor particles  has been
investigated in detail, for example, in \cite{kubis}.
$\Gamma$ can be expanded into functions which in turn are determined by angular momentum and parity of the corresponding meson:
\begin{eqnarray}
    \Gamma(p_4,\bp)&=&\sum\limits_\alpha g_\alpha(p_4,|\bp|) \,{\cal
    T}_\alpha(\bp) \label{spex} \: .
\end{eqnarray}
For pseudoscalar mesons there are 4 independent angular momentum functions ${\cal T}_\alpha(p)$ given by
\begin{eqnarray}
    {\cal T}_1(\bp)=\frac{\gamma_5}{\sqrt{16\pi}}, \quad {\cal
    T}_2(\bp)=\frac{\gamma_0\gamma_5}{\sqrt{16\pi}},\quad  {\cal
    T}_3(\bp)=\frac{1}{\sqrt{16\pi}}\frac{(\bp
    \bgam)}{|\bp|}\gamma_0\gamma_5, \quad {\cal
    T}_4(\bp)=\frac{1}{\sqrt{16\pi}}\frac{(\bp\bgam)}{|\bp|}\gamma_5 \: .
\label{nharms}
\end{eqnarray}
Our choice differs from the standard expansion \cite{Roberts}.
The main advantage of \eqref{nharms} is orthogonality $\int d\Omega_p Tr [{\cal T}_m(\bp) {\cal T}_n^+(\bp)]=\delta_{mn}$.

This property immediately allows to get a system of linear integral equations for the functions $g_{\alpha}(p_4,\bp)$.
As in the case of  the Schwinger-Dyson equation,  for these functions we employ also the  hyperspherical harmonics
\begin{equation}
    X_{kl}(\chi) = 2^l l! \sqrt{\frac{2}{\pi}} \sqrt{\frac{(k+1)(k-l)!}{(k+l+1)!}} \sin^l \chi C_{k-l}^{l+1}(\cos\chi)
\end{equation}
which separates the angular dependence
\begin{equation}
    g_i(p4,|\bp|) = \sum_j g^j_i(p) X_{j,0(1)}(\chi_p)
    \label{s0}
\end{equation}
and, hence, allows to perform the angular integration leaving us with a set of
coupled one-dimensional linear integral equations for the functions $g^k_i(p)$
\begin{eqnarray}
    g_i^k(p) = \int_0^\infty \frac {dk k^3}{4\pi^2} \sum A_{ij}^{km}(p,k)
    g_j^m(k)
    \label{int}
\end{eqnarray}
where $p$($k$) is the length of the Euclidean 4-momentum, $p=\sqrt{p_4^2+\bp^2}$.
The matrices $A_{ij}^{km}(p,k)$ can be calculated analytically.

The series  \eqref{s0} converges rather fast and in practice only  few
terms need to  be taken into account. Then by choosing a suitable method
of integration, e.g. Gaussian quadrature,
the system of equations (\ref{int}) can be written as
a system of homogeneous linear equations. Schematically, it can be written in the form
\begin{equation}
\label{bsem}
g=Kg \: .
\end{equation}
The condition
${\rm det}(K-1)=0$
is sufficient for the existence of a bound
state.
Hence, formally the zeros of the determinant ${\rm det}(K-1)$
determine the solution of the BS equation, including also excited states.
In our case,  for the sake of increasing the accuracy,  for the standard  Gaussian quadrature
 we use a  mapping
\begin{equation}
k= k_0 \frac{1+x}{1-x} \: ,
\end{equation}
where $k_0$ weights the integrand and $x \in [-1,1]$ is the integration variable (cf. Ref. \cite{our}).

Here an important moment is worth to be emphasized. Usually, when solving the BS equation for constituent
particles, i.e. for particles with constant masses, the resulting system of  partial equations is real.
In case of momentum dependent masses the
 Bethe-Salpeter equation becomes complex and requires the knowledge of the quark propagator for
 complex momenta $k_\pm$ (where  $k_\pm=\frac12 P\pm k$ are the momenta of quarks) and,
hence, requires to solve  the   Schwinger-Dyson equation also  for complex momenta.
For small meson masses (e.g.\ $M\approx500$ MeV) only a small energy region contributes to the
 integral in Eq. \eqref{int}, $k<1$ GeV.
In this case the propagator functions can be obtained by using the solution for real
 momenta, and   afterwards $A(p)$ and $B(p)$ are calculated at complex momenta from \eqref{ab} \cite{rob-1,Alkofer}.
For heavier states, $M>1$ GeV, the imaginary part of the quark momenta, Im $k_\pm$,
 becomes rather large and the integrand in \eqref{ab}
rapidly oscillates as a  function of $k$, hindering an accurate  computation of the integral.

\begin{figure} \begin{center}
\includegraphics[width=0.35
\textwidth]{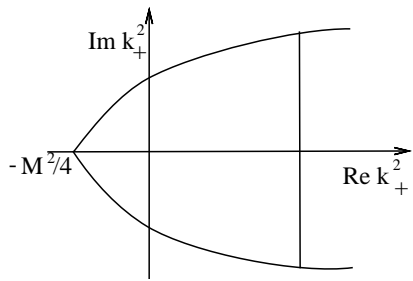}\end{center}
\caption{Integration contour  of (\ref{cauchy}).
Branches of parabola are determined by ${\rm Re}
k_{\pm}^2=-M^2/4-k^2$ and ${\rm Im} k_{\pm}=\pm Mk$. Since the integration
domain is infinite, in concrete calculations one shall restrict to some
large, but finite values of $k^2$; the  vertical line illustrates such a cut of the
infinite domain in practical calculations.
}
\label{fig2}
\end{figure}

To avoid this problem a witty trick has been suggested  in Ref. \cite{fisher}.
 It is based on the observation that,
as   seen from the  Schwinger-Dyson
equation,   all the  values of the momenta  $k_\pm^2$ are located within a domain limited by a parabola, as
shown in  Fig.~\ref{fig2}.

Then, due to the Cauchy's theorem, it suffices to know the solution along this parabola to be able to compute
it everywhere inside the corresponding domain\footnote{Many other methods may be employed
 to improve the accuracy  of numerical calculations, see e.g. Ref. \cite{ioak}.}:

\begin{eqnarray}
2\pi i A(z_1) = \oint dz \frac {A(z)}{z-z_1} \: .
\label{cauchy}\end{eqnarray}

Shifting the integration variable $l\rightarrow p-l$,  \eqref{ab} can be solved directly along this contour.
For large real part of the momentum the asymptotic form of $A$ and $B$ can be also used.
It should be noted that the described procedure of solving numerically
the  equations works quite well  if the integration domain is chosen in a reasonable way, i.e.
large enough to assure good asymptotics, but not too large, as the accuracy of the solution
decreases with increasing integration domain  (certainly, at a given Gaussian mesh).
Note also, that  asymptotic behaviors of the functions $A$ and $B$ is essential for heavier
quarks, see Fig.~\ref{fig1}.

\section{Numerical results}
\label{num}

\begin{figure} \begin{center}
\includegraphics[width=0.9
\textwidth]{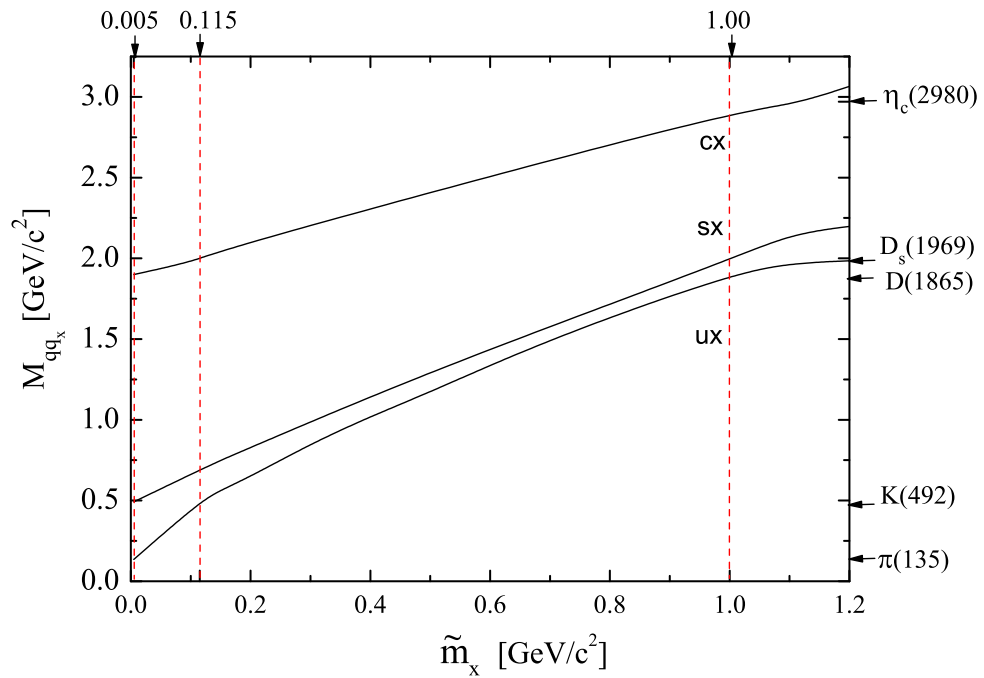}\end{center}
\caption{The bound state masses of a system $qq_x$ as a function of the bare  mass
$\tilde{m}_x$, the other quark being either a $u$ quark with
$\tilde{m}_u=0.005$ GeV (the curve labeled as "$ux$")
or a strange quark with $\tilde{m}_s=0.115$ GeV (the curve labeled as "$sx$") or a
charm quark with $\tilde{m}_c = 1.0$ GeV (the curve labeled as "$cx$").
The effective parameters are  $\omega = 0.5$ GeV and $D=16$ GeV$^{-2}$.
The masses of pseudoscalar $\pi$ $K$, $D$, $D_s$ and $\eta_c$ mesons according
to \cite{PDG} are indicated at the right side.
 The vertical dashed lines mark the selected bare quark masses for $u$, $s$, and $c$.}
\label{fig3}
\end{figure}

As an example of our numerical study we exhibit in Fig.~\ref{fig3}
the energy of the lowest bound states of a hypothetical meson $qq_x$ consisting of one given quark
$q$ with the mass known from the Schwinger-Dyson equation, bound with
a second quark $q_x$ for which the input bare mass  $\tilde{m}_x$  is let to  vary arbitrarily.
The corresponding effective parameters have been chosen as
$\omega = 0.5$ GeV and $D=16$ GeV$^{-2}$ \cite{Alkofer,Roberts} and the bare
masses for $q$ correspond to $(u,d,s)$ quarks,
q=u (with $\tilde{m}_u = 0.005$ GeV), q=s (with $\tilde{m}_s = 0.115$ GeV) and q=c (with $\tilde{m}_c = 1.0$ GeV).
This figure illustrates   the whole mass spectrum of pseudoscalar mesons with masses up to $ 3 \,\,GeV/c^2$. So, if
 the $q_x$ quark corresponds to a $c$ quark, then at the intersection of the vertical line
$\tilde{m}_x=m_c$ ($\approx 1 \,\, GeV/c^2$) with "$ux$" curve one obtains the $D$ -meson (with the quark contents a$uc$),
with the "$sx$" curve the $D_s$ meson and with the "$cx$" curve - the $\eta_c$ meson, respectively.
It is worth noting that the "$ux$" curve crosses the
$\tilde{m}_u=0.115$ GeV line roughly at the same value of $M_{qq_x}$ as the "$sx$" curve crosses
the $\tilde{m}_u=0.005$ GeV line,
thus proving a check  of consistency of the approach and, at the same time,
describing correctly the lowest pseudoscalar $us$ state corresponding to the K meson.
It can be seen that even without a fine tuning of $\tilde{m}_{s,c}$ the meson mass spectrum
is reproduced fairly well:  135 MeV ($\pi^0$ meson),
 497 MeV ($K$ meson), 1870 MeV ($D^\pm $ meson), 1970 MeV ($D_s^\pm $ meson)
 and 2980 MeV  ($\eta_c$ meson).

In nature, a pseudoscalar $ s \bar s $ meson with simple quark structure does not exist.
However, our  "$sx$"  curve intersects with the bare  mass  of the  $s$-quark, i.e. the approach predicts
an existence of $s\bar s$ meson around 600 MeV.   This can be considered
as a "ghost" state, or as check of   consistency of the approach.
Apart from this circumstance, it is amazing  that the simple two-parameter ansatz of the gluon propagator in the
IR region together with the Schwinger-Dyson equation in rainbow approximation delivers such a
quark propagator which, via the Bethe-Salpeter equation in rainbow-ladder approximation,
results in
such a nice description of pseudoscalar $\pi$, $K$, $D$, $D_s$ and $\eta_c$ states.

Clearly, a special parameterization of the gluon propagators with two adjusted quantities
and the self consistent determination of three bare quark masses serve as input for obtaining
finally five pseudoscalar meson masses.
The straight forward extension to the scalar, vector and axialvector mesons
without further adjustments will provide the ultimate test for the subtleties
of the numerical implementation of the employed theoretical scheme and the inherent approximations.
Work along this line is in progress.

Masses of excited states, which correspond to radial excitations of the considered
mesons, can be evaluated in a similar way, i.e.\ by finding of the
next zeros of the determinant (\ref{bsem}). As   our analysis shows,
 the determinant changes monotonously with increasing mass.
So that, naively, one would not expect a bound state of a $ u \bar u$ system with mass larger than 800 MeV,
since even at zero momenta, the maximum constituent quark mass is around 400 MeV, cf. Fig.~\ref{fig1}.
However,  in the deeply non-Euclidean  domain, which corresponds to $k_+^2\simeq -M^2/4$ (Fig.~\ref{fig2}),
the dynamical quark masses, contrubiting to the BS equation,
  increase till 600-700 MeV depending on $M$. Hence, the Schwinger-Dyson equation allows to understand
  the formation mechanism of such bound states which, from the constituent quark model point of view,
    can not be even predicted a priori.
   The mass of the first excited $u\bar u$ state is
   found  to be 1080 MeV, i.e. significantly  above the maximum mass from the Schwinger-Dyson equation alone.
   Analogously,  for the $ c \bar u$ system, the first excited state is found to be around  2530 MeV, which is in a good
   agreement   with data.
   Similar results have been obtained in other groups, see Ref. \cite{krassnigg1}.

Note that in solving the BS equation for the mass spectrum of mesons we obtain also the partial BS amplitudes which
can be used in calculations of various dynamical observables, such as the meson life-time,
transition form factors, the
dependence of the meson widths up on temperature of an ambient medium etc. Such
investigations are  in progress and results
will be reported elsewhere.

\section{Conclusion}
\label{concl}

The method of solving the Schwinger-Dyson equation in rainbow-ladder approximation in  Euclidean
space by using the hyperspherical harmonics basis is proposed.
The obtained numerical solutions are then used to solve the Bethe-Salpeter equation for the meson
mass spectrum in a large interval of meson masses.
In solving the Bethe-Salpeter equation a new  set of basis functions has been
used which allows a further  easy decomposition of the Bethe-Salpeter vertex functions
 into hyperspherical harmonics basis

The obtained mass spectrum  for pseudoscalar mesons
in  a wide range, ranging from  pions  to  $\eta_c$ mesons,
 is in a good agreement with experimental data.
Excited states were considered as well and found to be also  in a good  agreement with experimental data and
with calculations by other groups. By solving the Bethe-Salpeter  equation, the corresponding
partial wave functions are also obtained which will allow, in future, to calculate a variety of observables
related to  physical programmes at, e.g. FAIR.

\section*{Acknowledgments}
This work was supported in part by the Heisenberg - Landau program
of the JINR - FRG collaboration, GSI-FE and BMBF 06DR9059. Calculations were partially
performed on the Caspur facilities under the Standard HPC 2010 grant "SRCnuc".~

\end{document}